\documentclass[12pt]{paper}
\usepackage{amsmath}
\usepackage{amssymb}
\usepackage{amsfonts}
\usepackage[ansinew]{inputenc}
\usepackage{color}
\newcommand{\be}{\begin{eqnarray}}
\newcommand{\ee}{\end{eqnarray}}
\newcommand{\noi}{\noindent}
\newcommand{\dep}{\partial}

\newcommand{\edt}{elasticity difference tensor}

\begin{document}

\title{General spherically symmetric elastic stars in Relativity}
\author{I. Brito$^\ast$, J. Carot$^\natural$, E.G.L.R. Vaz$^\ast$ }

\maketitle

$^\natural$Departament de Física, Universitat de les Illes
Balears,\\ Cra Valldemossa pk 7.5, E-07122 Palma de Mallorca, Spain\\
$^\ast$Departamento de Matemática para a Ciência e Tecnologia,\\
Universidade do Minho, Guimar\~{a}es, Portugal

\begin{abstract}

The relativistic theory of elasticity is reviewed within the
spherically symmetric context with a view towards the modeling of
star interiors possessing elastic properties such as the ones
expected in neutron stars. Emphasis is placed on generality in the
main sections of the paper, and the results are then applied to
specific examples. Along the way, a few general results for
spacetimes admitting isometries are deduced, and their consequences
are fully exploited in the case of spherical symmetry relating them
next to the the case in which the material content of the spacetime
is some elastic material. This paper extends and generalizes  the
pioneering work by Magli and Kijowski \cite{Magli1}, Magli \cite{Magli2}
and \cite{M1}, and complements, in a sense, that  by Karlovini
and Samuelsson in their interesting series of papers \cite{KS1},
\cite{KS2} and \cite{KS3}.

\end{abstract}

\section{Introduction}
\label{Introduction}

The interest of a relativistic theory of elasticity is twofold; on
the one hand there is its purely theoretical interest, namely that
of providing a relativistic extension of a well-known (and very
fruitful) classical theory; on the other hand  and  on theoretical
grounds, it is expected that neutron stars possess a solid crust
with elastic properties which may help explaining certain
observational issues (see \cite{KS1} for a thorough account of these
and other related features). Further, anisotropy in pressures is a
phenomenon occurring in many situations of equilibrium which are of
interest in astrophysics and whose corresponding dynamics has been
thoroughly studied (see for instance \cite{Herrera} and references
therein); the assumed point of view however has been an heuristic
one, without providing mechanisms explaining how the anisotropy in
pressures may arise and using instead \emph{ad hoc} assumptions.
Magli and Kijowski \cite{Magli1} and Magli \cite{M1} have shown that
in spherical symmetry, the anisotropy in pressures arises quite
naturally as the relativistic extension of the classical
(non-relativistic) non-isotropic stress in elasticity theory. See
also \cite{Park} for an excellent study of the static case in
spherical symmetry, where existence theorems for regular solutions
near the center are proven under rather mild, physically meaningful,
hypotheses. Beig and Schmidt \cite{BS1} have shown that, in general,
the field equations for elastic matter can be cast into a
first-order symmetric hyperbolic system and that as a consequence,
local-in-time existence and uniqueness theorems may be obtained
under various circumstances.

The aim of this paper is to extend and generalize the work presented
in \cite{Magli1} and \cite{M1} as well as to set up a set of
mathematical tools and equations that may facilitate the obtention
of exact solutions to Einstein´s Field Equations (EFE) describing
the interior of elastic materials and satisfying the Dominant Energy
Condition (DEC).

The paper is organized as follows: in the next section we provide a
brief account of the theory of relativistic elasticity, much along
the lines followed in \cite{Magli1}, \cite{M1} and \cite{KS1}, but
we shall also include some comments on the relationship between the
isometries in the material space and in the spacetime. Most of the
results in that section are well known and could be found in the
above references, but we are still including them in order to set up
the notation which will be followed in the remainder of the paper as
well as for making the present paper more self-contained.  Section 3
contains a digression on spherically symmetric spacetimes and the
restrictions that such an assumption imposes on the physics in these
spacetimes, which we then apply it to the case of elastic materials.
In section 4 EFEs are obtained for the general case and some
particular cases are commented upon. In section 5 we
analyze in detail the case of shearfree solutions, paying special attention to the fulfilment of the dominant enrgy condition as well as to the necessary and sufficient conditions that must be satisfied for an equation of state to be admitted; finally we p
 resent a few selected examples; these include the analysis of the
\edt \ for the nonstatic case, most along the lines followed in a
previous paper by two of the authors \cite{VB}.

\section{Relativistic elasticity revisited}

Let $(M,g)$ be a spacetime, $M$ then being a 4-dimensional
Hausdorff, simply connected manifold of class $C^2$ at least, and
$g$ a Lorentz metric of signature $(-,+,+,+)$. The \emph{material
space} $X$ is a 3-dimensional  manifold endowed with a Riemannian
metric $\gamma$, the \emph{material metric}; points in $X$ can then
be thought of as the particles of which the material is made of.
Coordinates in $M$ will be denoted as $x^a$ for $a=0,1,2,3$, and
coordinates in $X$ as $y^A$, $A=1,2,3$. The material metric $\gamma$
is not a dynamical quantity of the theory, but it is frozen in the
material, and it roughly describes distances between neighboring
particles in the relaxed state of the material.

The spacetime configuration of the material is said to be completely
specified whenever a submersion $\psi: M \rightarrow X$ is given; if
one chooses coordinate charts in $M$ and $X$ as above, the
coordinate representative of $\psi$ is given by three fields
$$  y^A = y^A(x^b), \qquad A=1,2,3$$ and the physical laws
describing the mechanical properties of the material can then be
expressed in terms of a hyperbolic second order system of PDE. The
differential map $\psi_*: T_pM \rightarrow T_{\psi(p)} X $ is then
represented in the above charts by the rank 3 matrix
$$ \left( y^A_{\; b}\right)_p, \qquad  y^A_{\; b}=\frac{\dep y^A}{\dep
x^b} \qquad A=1,2,3, \;\; b=0,1,2,3$$ which is sometimes called
\emph{relativistic deformation gradient}. Since $\psi_*$ has maximal
rank 3, its kernel is spanned at each point by a single timelike
vector which we may take as normalized to unity, the resulting
vector field, say $\vec u = u^a\dep_a$, satisfies then
$$y^A_{\; b} u^b = 0, \quad u^au_a = -1, \;\; u^0 >0$$ the last
condition stating that we choose it future oriented; $\vec u$  is
called the \emph{velocity field of the matter}, and in the above
picture in which the points in $X$ are material points, it turns out
that the spacetime manifold $M$ (or, to be more precise, an open
submanifold of it) is then made up by the worldlines of the material
particles, whose tangent vector is precisely $\vec u$.

The material space is said to be in a \emph{locally relaxed state}
at an event $p\in M$ if, at $p$, it holds $k_{ab}\equiv (\psi^*
\gamma)_{ab} = h_{ab}$ where $h_{ab} = g_{ab} + u_au_b$. Otherwise,
it is said to be \emph{strained}, and a measurement of the
difference between $k_{ab}$ and $h_{ab}$ is the  \emph{strain},
whose definition varies in the literature; thus, it can be defined
simply as $S_{ab} = -\frac12(k_{ab} - h_{ab})=-\frac12(k_{ab}-u_au_b
- g_{ab})$. We shall follow instead the convention in \cite{M1} and
use \be\label{2.0.0} K_{ab} \equiv k_{ab}-u_au_b\ee Notice that
$K^a_{\, b} u^b = u^a$, and therefore one of its eigenvalues is $1$.
Definitions using the logarithm of the above tensor\footnote{With
one index raised, thus one has a linear operator which turns out to
be positive and self-adjoint, its logarithm being then well-defined}
also appear in the literature as that allows simple interpretations
of the associated algebraic invariants, see e.g. \cite{Magli1} and
\cite{Park}.

The strain tensor determines the elastic energy stored in an
infinitesimal volume element of the material space (or energy per
particle), hence that energy will be a scalar function of $K_{ab}$.
This function is called \emph{constitutive equation} of the
material, and its specification amounts to the specification of the
material. We shall represent it as $v = v(I_1,I_2,I_3)$, where
$I_1,I_2,I_3$ are any suitably chosen set of scalar
invariants\footnote{Recall that one of the eigenvalues is 1,
therefore, there exist three other scalars (in particular they could
be chosen as the remaining eigenvalues) characterizing $K^a_{\, b}$
completely along with its eigenvectors.} associated with and
characterizing $K_{ab}$ completely. Following \cite{M1} we shall
choose
\begin{equation}\begin{split}\label{I1.0} {I_{1}}&=\frac{1}{2}\left(\text{Tr} {K}-4\right)\\
 {I_{2}}&=\frac{1}{4}\left[\text{Tr} {K}^{2}-\left(\text{Tr} {K}\right)^{2}\right]+3\\
 {I_{3}}&=\frac{1}{2}\left(\text{det} {K}-1\right),\end{split}\end{equation}

Notice that for $K_{ab} = g_{ab}$ (equivalently $k_{ab} = h_{ab}$)
the strain tensor $S_{ab}$ is zero, that is: the induced metric on
the rest frame of an observer moving with four-velocity $\vec u$,
$h$, coincides with the material metric $\gamma$ (its pull-back by
$\psi$) describing the relaxed state of the material; thus it makes
sense to have zero elastic energy stored. It is immediate to check
from the above expressions that in this case one has $I_1=I_2=I_3 =
0$.

The energy density $\rho$ will then be the particle number density
$\epsilon$ times the constitutive equation, that is \be \label{rho0}
\rho = \epsilon v(I_1,I_2,I_3) = \epsilon_0 \sqrt{\det K}\,
v(I_1,I_2,I_3)\ee where $\epsilon_0$ is the particle number density
as measured in the material space, or rather, with respect to the
volume form associated with $k_{ab} = (\psi^* \gamma)_{ab}$, and
$\epsilon$ is that with respect to $h_{ab}$; see \cite{KJ1} for a
proof of the above equation. In some references (e.g. \cite{M1}),
the names $\rho$ and $\epsilon$ are exchanged and the density
measured w.r.t. $k_{ab} = (\psi^* \gamma)_{ab}$ ($\epsilon_0$ in our
notation) is then called ``density of the relaxed material'' (see
the above comments on the meaning of $\gamma$), whereas that
measured w.r.t. $h_{ab}$ is referred to as the ``density in the rest
frame''.

We next turn our attention towards the energy-momentum tensor of an
elastic material. Before proceeding, it will be useful to recall
that any symmetric, second order covariant tensor field may be
decomposed with respect to a timelike unit vector field $\vec v, \;
v^av_a =-1$ as follows: \be\label{2.0}T_{ab}=\rho v_a v_b+ph_{ab}+
P_{ab}+v_aq_b+q_av_b \ee where $h_{ab} = g_{ab} + v_a v_b$, $P_{ab}=
h^m_{\, a}h^n_{\, b}(T_{mn}-3p h_{mn})$, $q_a =-(T_{ab}v^b + \rho
v_a)$, $\rho =T_{ab}v^av^b$, $p= \frac{1}{3} h^{ab}T_{ab}$. From the
definitions of these variables it readily follows $$h_{ab}v^b =0,
\quad P_{ab} v^b = g^{ab} P_{ab} =0, \quad \mathrm{and}\quad q^av_a
= 0.$$ In the case that $T_{ab}$ represents the energy-momentum
tensor of some material distribution, $\rho, \, p, \,P_{ab}, \, q^a$
are respectively the energy density, isotropic pressure, anisotropic
pressure tensor and heat flow that a family of observers moving with
four-velocity $\vec v$ would measure at every point in the
spacetime.

In the case of elastic matter, it can be seen using the standard
variational principle for the Lagrangian density $\Lambda =
\sqrt{-g} \rho$ (see for instance \cite{Magli1} or \cite{KS1} and
the beginning of section 5 for further details) that the
energy-momentum takes the form, when decomposed with respect to
$\vec u$, the velocity of the matter:
\begin{equation}\label{2.1}
    T_{ab}=\rho u_a u_b+ph_{ab}+P_{ab}
\end{equation}
where all the definitions are the same as the ones given above
substituting $\vec u$ for $\vec v$, i.e.: $h_{ab} = g_{ab} + u_a
u_b$, $P_{ab}= h^m_{\, a}h^n_{\, b}(T_{mn}-3p h_{mn})$, $\rho
=T_{ab}u^au^b$, $p= \frac{1}{3} h^{ab}T_{ab}$ and they satisfy
$h_{ab}u^b =0, \;P_{ab} u^b = g^{ab} P_{ab} =0$; thus in particular
one gets $q^a = 0$ and the resulting tensor is of the diagonal Segre
type $\{1, 111\}$ or any of its degeneracies, $\vec u$ being its
(unit) timelike eigenvector (see \cite{KSMH}).

This means that an orthonormal tetrad exists $\{u_a, x_a, y_a,
z_a\}$ (with $u_au^a=-1$, $x^ax_a = y^ay_a =z^az_a =+ 1$ and the
mixed products zero) with respect to which $T_{ab}$ may be written
as
\begin{eqnarray}
    T_{ab}=\rho u_a u_b+p_1 x_{a}x_{b}+p_2y_{a}y_{b}+ p_3z_{a}z_{b},
    \quad p =\frac{1}{3} (p_1 + p_2 + p_3),\nonumber \\ h_{ab} = x_{a}x_{b}+y_{a}y_{b}+
    z_{a}z_{b},  \quad \mathrm{etc.}\qquad\qquad\qquad\qquad\label{2.2}
\end{eqnarray}
It is interesting to mention that the Dominant Energy Condition
(DEC), see for instance \cite{KSMH}, is fulfilled if and only if
\begin{equation}\label{2.3}
    \rho \geq 0, \qquad  \vert p_A \vert \leq \rho, \qquad A=1,2,3.
\end{equation}

\section{On symmetries and their consequences on physics}

Let $(M,g)$ admit  a Killing Vector (KV) $\vec \xi$, i.e.: in any
coordinate chart $x^a$, $\mathcal{L}_{\vec \xi} \ g_{ab} =0$. It is
then immediate to show that $\mathcal{L}_{\vec \xi}\ R_{ab}
=\mathcal{L}_{\vec \xi} \ G_{ab}= \mathcal{L}_{\vec \xi} \
R^a_{\;bcd}=0$, etc. and, from EFEs it then follows that
$\mathcal{L}_{\vec \xi}\ T_{ab} =0$ where $T$ denotes the
energy-momentum tensor describing the material in the spacetime.

It is easy to show that, if $\vec v$ is a non-degenerate unit (i.e.:
$v^av_a = \epsilon$ with $\epsilon = \pm 1$) eigenvector of $T_{ab}$
with corresponding eigenvalue $\lambda$, then \be\label{simetria1}
\mathcal{L}_{\vec \xi} \ \lambda = \mathcal{L}_{\vec \xi}\ v^a = 0.
\ee We next include a short proof of this.

\begin{quotation}

\noi (i) Taking the Lie derivative of  $T_{ab}v^b$, with respect to
$\vec \xi$ and since we are assuming that $T_{ab}v^b  = \lambda v_a$
one gets \be \mathcal{L}_{\vec \xi} \ (T_{ab}\;
v^b)=\mathcal{L}_{\vec \xi} \ (\lambda \; v_a) = \lambda\;
\mathcal{L}_{\vec \xi}\ v_a\;+ \; v_a  \mathcal{L}_{\vec \xi}\
\lambda. \label{3-1} \ee On the other hand, \be \mathcal{L}_{\vec
\xi}\ (T_{ab}\ v^b)=T_{ab}\ \mathcal{L}_{\vec \xi}\ v^b.
\label{3-2}\ee Equating \eqref{3-1} and \eqref{3-2} and then
contracting with $v^a$, yields \be \lambda \ v_b (\mathcal{L}_{\vec
\xi}\ v^b)=\lambda \ v^a \mathcal{L}_{\vec \xi}\ v_a\; +\;
\mathcal{L}_{\vec \xi}\ \lambda. \ee Therefore $\mathcal{L}_{\vec
\xi} \ \lambda =0,$ since as $\vec\xi$ is a KV $v_a
\mathcal{L}_{\vec \xi}\ v^a  =  v^a \mathcal{L}_{\vec \xi}\ v_a $.
\vspace{0.3cm}

\noi (ii) Substituting this result into \eqref{3-1} and using
\eqref{3-2} one obtains $T_{ab}\mathcal{L}_{\vec \xi}
 v^b= \lambda \; \mathcal{L}_{\vec \xi} v_a.$ Therefore, $v^a$ and the vector $w^a \equiv \mathcal{L}_{\vec
 \xi} v^a$ are eigenvectors of $T_{ab}$ associated with the same
 eigenvalue. Since this is non-degenerate, $\vec v$ and $\vec w$
 have to be proportional, i.e. $\mathcal{L}_{\vec \xi} v^a= \alpha
 v^a$, for some real value $\alpha$, however this $\alpha$ must be zero
 as $\vec v$ is unit and $\vec \xi$ is a KV, indeed $$ 0 = \mathcal{L}_{\vec \xi}
 v^a v_a = 2 v_a\mathcal{L}_{\vec \xi}
 v^a = 2\alpha v^av_a, \qquad \mathrm{hence}\quad \alpha=0.    $$

\end{quotation}

Under the hypothesis that $\vec \xi$ is a Killing vector, and on
account of the above considerations the following conditions hold in
the case that $T_{ab}$ represents elastic matter and is therefore of
the form (\ref{2.1})

$$\mathcal{L}_{\vec \xi} g_{ab}=0\; \Rightarrow \; \mathcal{L}_{\vec \xi} \rho=0,\;
\mathcal{L}_{\vec \xi} u_a =0,\;  \mathcal{L}_{\vec
 \xi} h_{ab} =0,\;  \mathcal{L}_{\vec \xi} P_{ab}=0 ,\;
 \mathcal{L}_{\vec \xi} p=0.$$

The first two are just the specialization of the above comments to
the case $\vec v =\vec u$ and $\lambda =\rho$. The vanishing of
$\mathcal{L}_{\vec \xi} h_{ab} $ follows then from the vanishing of
the Lie derivative of the metric and that of $u_a$; notice that
$\mathcal{L}_{\vec \xi} h^{ab}=0 $ as well. Next, since $p=
\frac{1}{3}h^{ab}T_{ab}$  it also follows that its Lie derivative
with respect to the KV vanishes as those of $T_{ab}$ and $h^{ab}$
do, and the vanishing of $P_{ab}$ follows then immediately from the
expression (\ref{2.1}) and the vanishing of the Lie derivatives of
all the other terms. Therefore we have shown that matter
$4-$velocity, pressure, density, anisotropic tensor all stay
invariant along the Killing vectors of the space-time, together with
the projection tensor $h_{ab}= g_{ab}+u_a
 u_b $. It is interesting to notice that, for a general
energy momentum tensor such as (\ref{2.0}), and if one assumes that
$\mathcal{L}_{\vec \xi} u_{a}=0$, it also follows that
$\mathcal{L}_{\vec \xi} q_a =0$; but in this case that assumption
has to be made, as $\vec u$ is no longer an eigenvector of the
energy-momentum tensor.

Similar conclusions (although not the same) can be drawn when $\vec
\xi$ is a proper homothetic vector field; i.e.: $\mathcal{L}_{\vec
\xi}\ g_{ab} = 2kg_{ab}$, with $k\neq 0$; in which case one also has
$\mathcal{L}_{\vec \xi} \ T_{ab} = 0$ but then, for instance,
$\mathcal{L}_{\vec \xi}\ u_{a} =-ku_a$ and also $\mathcal{L}_{\vec
\xi} \ \rho = k\rho$, etc.

In this paper we shall be concerned with the case of elastic
materials in spherically symmetric spacetimes. We next explore the
consequences that the existence of symmetries has on the material
content of a spacetime. Most of the developments following are well
known although disperse in the literature, we collect them here for
the sake of completeness.

As it is well known, for a spherically symmetric spacetime,
coordinates \; $x^a=t,r,\theta, \phi$ exist (and are non-unique)
such that the line element can be written as \be ds^2= -a(r,t) dt^2
+ b(r,t) dr^2+ r^2 d \theta^2 + r^2 \sin^2 \theta d \phi^2
\label{metrica1} \ee with $a$ and $b$ positive and independent of
$\theta$ and $\phi$. This metric possesses three Killing vectors,
namely $\vec{\xi_1}=-\cos \phi\;
\partial_{\theta}+\cot \theta \sin \phi \;
\partial_{\phi} ,$ \; $ \vec{\xi_2}=\partial_{\phi} $ \; and \;
$ \vec{\xi_3}=-\sin\phi \;
\partial_{\theta}-\cot \theta \cos \phi \; \partial_{\phi} $ which
generate the 3-dimensional Lie algebra $so(3)$.

To start with, we show that any timelike vector field $\vec v$ that
remains invariant along the three Killing vectors is necessarily of
the form $$\vec v= v^t (t,r) \;
\partial_t + v^r (t,r) \;
\partial_r.$$

Using $\mathcal{L}_{\vec \xi_2}  v^a =0 $ for $a=0,1,2,3$ we
conclude that all the components $v^a$ are independent of $\phi$.
Then, the expression
$$\mathcal{L}_{\vec \xi_1}  v^a = \xi_1^c \; v^a_{,c}-v^c \;
\xi_{1,c}^a,$$  for $a=0,1 \ $ gives that $v^0$ and $v^1$ are also
independent of $\theta$ and for $a=2,3$ yields $v^2=v^3=0.$

It should be noticed that it always exists a coordinate
transformation taking $t,r$ into $t', r'$  such that $\vec v =
v^{t'} (t',r') \; \partial_{t'}$ and the metric \eqref{metrica1}
reads then \be ds^2= -a(r,t) dt^2 + b(r,t) dr^2+ Y^2(r,t)\left( d
\theta^2 + \sin^2 \theta d \phi^2\right) \label{metrica2} \ee where
primes have been dropped for convenience. These new coordinates are
the so called \emph{comoving coordinates}; one can see this either
by direct computation, showing that one such coordinate change is
always possible, or else, by showing first that any vector field
such as $\vec v$ above is always hypersurface orthogonal, that is:
$\omega_{ab} = v_{[a;b]} + \dot v_{[a}v_{b]} = 0$, then it follows
that  $v_a \propto \dep_a t'$ for some function $t'$, choosing this
function as the new time coordinate, one can readily show the above
result.

Next, for any symmetric, second order tensor $P_{ab}$, which is traceless, invariant under the three KVs above, and orthogonal to a vector such as $\vec v$ above (i.e.: timelike spherically
symmetric), that is:
\begin{enumerate}

\item[(i)] $P_{ab}v^b=0$

\item[(ii)] $g^{ab}P_{ab}=0$

\item[(iii)] $\mathcal{L}_{\vec \xi_A} P_{ab}=0,$ for $A=1,2,3$
\end{enumerate}

it follows that $P_{ab}$ is proportional to the shear tensor of
$\vec v$ whenever the latter is non-zero, namely: \be\label{3.1.1}
P_{ab} \propto \sigma_{ab}, \qquad
\sigma_{ab}=v_{(a;b)}+\dot{v}_{(a} v_{b)}-\frac13 \theta h_{ab}\ee
where $\theta= v^a_{;a} $ is the expansion, $\dot{v}_a=v_{a;b}v^b$
is the acceleration and round brackets denote symmetrization as
usual.

This can be proven easily by making use of the comoving coordinate
system referred to above, and imposing the various conditions (i -
iii); also, a more general proof is possible in the context of
warped spacetimes (of which the spherically symmetric ones are
special instances), see \cite{CarotNunez}.

In the comoving coordinate system above, one can see by direct
computation that the shear $\sigma_{ab}$ is \be\label{shear1}
\sigma_{ab}= \mathrm{diag}\left(0,\frac13 \frac{(b_tY -2 Y_t
b)}{\sqrt{a}Y}, -\frac16\frac{Y(b_tY -2 Y_t b)}{\sqrt{a}Y},
-\frac16\frac{Y(b_tY -2 Y_t b)}{\sqrt{a}Y}\sin^2 \theta \right) \ee
and therefore this field is shearfree if and only if
$$ b(r,t) = F^2(r)Y^2(r,t),$$
in which case it is always possible, by means of an obvious
redefinition of the coordinate $r$,  bring the metric to the form
\be ds^2= -a(r,t) dt^2 + Y^2(r,t)\left( dr^2+  d \theta^2 + \sin^2
\theta d \phi^2\right). \label{metrica3} \ee

For the class of spacetimes we shall be interested in, namely
elastic, spherically symmetric, all the above apply for $\vec u$
(the velocity of matter), as it is indeed the unit timelike
eigenvector of $T_{ab}$ given by \eqref{2.1}, and $P_{ab}$ the
anisotropic pressure tensor  since, as  discussed previously, it is
invariant under the KVs the spacetime possesses and is also
traceless and orthogonal to $\vec u$, therefore we have that,
whenever the shear of $\vec u$ is non-zero \be\label{3.1.2} P_{ab} =
2\lambda \sigma_{ab}, \qquad \sigma_{ab}=u_{(a;b)}+\dot{u}_{(a}
u_{b)}-\frac13 \theta h_{ab}\ee where $\theta=u^a_{;a} $ and
$\dot{u}_a=u_{a;b}u^b$, and $\lambda=\lambda(t,r)$ is some function,
therefore, for the generic (non shearfree) case, it is always
possible to treat, at least formally\footnote{There is a further
requirement for a fully physically meaningful interpretation as a
viscous fluid, namely that $\lambda <0$ in which case the
kinematical viscosity would be $\eta = -\lambda >0$.}, the elastic
material as a viscous fluid with zero heat flow. This interpretation
would indeed break down in the case in which $\vec u$ is  shearfree.

\section{Elasticity in spherical symmetry}

Let us now consider in more detail the problem of elasticity in a
spherically symmetric spacetime $(M,\bar g)$ with associated
material space $(X,\bar \gamma)$.

The results given in this section generalize those in \cite{M1} in
the sense that here we consider a non flat material metric $\bar
\gamma$, while, when referring to quantities and results in
\cite{M1}, we shall use non-barred quantities (hence the bars on the
spacetime metric and the material metric in our notation).

Recalling the notation and results in section 2, we shall demand
that the submersion $\psi: M\longrightarrow X$ preserves the KVs,
that is: $\psi_*(\vec \xi_A) = \vec\eta_A$ are also KVs on $X$.

This implies that the metric $\bar\gamma$ is also spherically
symmetric and therefore coordinates $y^A =
(y,\tilde\theta,\tilde\phi )$ exist with $y=y(t,r)$,
$\tilde{\theta}=\theta$ and $\tilde{\phi}=\phi$, and are such that
$\vec\eta_A=\vec\xi_A$ are KVs of the metric $\bar\gamma$. Thus, the
line elements of $\bar g$ and $\bar\gamma$ may be written as:
\be\label{4.1} d\bar
s^{2}=-\bar{a}(t,r)dt^{2}+\bar{b}(t,r)dr^{2}+r^{2}d\theta^{2}+r^{2}sin^{2}\theta
d\phi^{2}\ee
 \be\label{4.2}
d\bar\Sigma^{2}=f^{2}(y)(dy^{2}+y^{2}d{\theta}^{2}+y^{2}sin^{2}{\theta}d{\phi}^{2}),
\ee Notice that this last expression is completely general, as any
3-dimensional spherically symmetric metric is necessarily
conformally flat, as it is immediate to show.

The results in  \cite{M1} correspond to $f(y)=1$, and the relation
between $\bar{\gamma}$ and the flat material metric $\gamma$ used in
\cite{M1} is given by \begin{equation}\label{M14}
\bar{\gamma}_{AB}=f^{2}(y)\gamma_{AB}.\end{equation}

Next, attention should be payed to the canonical definition of the
energy-momentum tensor used by \cite{M1}: \be
{T}^{a}_{b}=\frac{1}{\sqrt{-g}}\left(\frac{\partial\Lambda}{\partial
y^{A}_{a}}y^{A}_{b}-\delta^{a}_{b}\Lambda\right).\ee As shown in
\cite{KM1}, this canonical definition of the energy-momentum tensor
coincides with the symmetric definition of the energy-momentum
tensor, used by other authors, up to a sign, which is a particular
case of the general Belinfante-Rosenfeld theorem \cite{Bel},
\cite{Ros}.

Denoting by $\bar k$ the pull-back by $\psi$ of the material metric
$\bar\gamma$, that is: $\bar k = \psi^*(\bar\gamma)$, one has:
\begin{equation*}
\begin{split}
\bar{k}^{a}_{b}
&=\bar{g}^{ac}\bar{k}_{cb}=\bar{g}^{ac}\bar{\gamma}_{CB}y^{C}_{c}y^{B}_{b}=f^{2}(y)\bar{g}^{ac}\gamma_{CB}
y^{C}_{c}y^{B}_{b}=f^{2}(y)\bar{g}^{ac}{k}_{cb}\\
&=f^{2}(y)\bar{g}^{ac}[\dot{y}^{2}\delta^{0}_{c}\delta^{0}_{b}+
\dot{y}y'(\delta^{0}_{c}\delta^{1}_{b}+\delta^{0}_{b}\delta^{1}_{c})
+y'^{2}\delta^{1}_{b}\delta^{1}_{c}+y^{2}\delta^{2}_{b}\delta^{2}_{c}+y^{2}sin^{2}\theta
\delta^{3}_{b}\delta^{3}_{c}],
\end{split}
\end{equation*}
or
%
\begin{equation}\bar{k}^{a}_{b} =
\left(\begin{array}{cccc}-f^{2}(y)(\dot{y}^{2}/\bar{a})&-f^{2}(y)(\dot{y}y'/\bar{a})&0&0\\
f^{2}(y)(\dot{y}y'/\bar{b})&f^{2}(y)(y'^{2}/\bar{b})&0&0\\0&0&f^{2}(y)y^{2}/r^{2}&0\\0&0&0&f^{2}(y)y^{2}/r^{2}\end{array}
\right), \label{22}\end{equation} where a dot indicates a derivative with
respect to $t$ and a prime a derivative with respect to $r$.

The velocity field of the matter, defined by the conditions
$\bar{u}^{a}y^{A}_{a}=0$, $\bar{g}_{ab}\bar{u}^{a}\bar{u}^{b}=-1$
and $\bar{u}^{0}>0$, can be expressed as \be
\bar{u}^{a}=\frac{\bar{\Gamma}}{\sqrt{\bar{a}}}\left(1,-\frac{\dot{y}}{y'},0,0\right),
\ee where
\begin{equation}\label{gamma}
\bar{\Gamma}\equiv
\left(1-\frac{\bar{b}}{\bar{a}}\left(\frac{\dot{y}}{y'}\right)^{2}\right)^{-\frac{1}{2}}.
\end{equation}
Therefore the projection tensor is
\begin{equation}\bar{h}^{a}_{b}=\delta^{a}_{b}+\bar{u}^{a}\bar{u}_{b}= \left(\begin{array}{cccc}1-\bar{\Gamma}^{2}&
-\bar{\Gamma}^{2}(\bar{b}\dot{y}/\bar{a}y')&0&0\\
\bar{\Gamma}^{2}(\dot{y}/y')&1+\bar{\Gamma}^{2}(\bar{b}/\bar{a})(\dot{y}/y')^{2}&0&0\\0&0&1&0\\0&0&0&1\end{array}
\right).
\end{equation}
We will use an orthonormal tetrad and write the metric as
$\bar{g}_{ab}=-\bar{u}_{a}\bar{u}_{b}+\bar{x}_{a}\bar{x}_{b}+\bar{y}_{a}\bar{y}_{b}+
\bar{z}_{a}\bar{z}_{b},$ such that:
\begin{flalign*}
\bar{u}^{a}&
=\left(\frac{\bar{\Gamma}}{\sqrt{\bar{a}}},-\frac{\dot{y}}{y'}\frac{\bar{\Gamma}}{\sqrt{\bar{a}}},0,0\right)&
\bar{u}_{a}& =
\left(-\sqrt{\bar{a}}\bar{\Gamma},-\frac{\bar{b}}{\sqrt{\bar{a}}}\frac{\dot{y}}{y'}\bar{\Gamma},0,0\right)\\
\bar{x}^{a}&
=\left(-\frac{\sqrt{\bar{b}}}{\bar{a}}\frac{\dot{y}}{y'}\bar{\Gamma},\frac{\bar{\Gamma}}{\sqrt{\bar{b}}},0,0\right)&
\bar{x}_{a}&
=\left(\frac{\dot{y}}{y'}\sqrt{\bar{b}}\bar{\Gamma},\sqrt{\bar{b}}\bar{\Gamma},0,0\right)\\
\bar{y}^{a}& =\left(0,0,\frac{1}{r},0\right)& \bar{y}_{a}& =\left(0,0,r,0\right)\\
\bar{z}^{a}& =\left(0,0,0,\frac{1}{r\sin \theta}\right)&
\bar{z}_{a}& =\left(0,0,0,r\sin \theta\right),
\end{flalign*}
where $\displaystyle{\bar{\Gamma}}$ is the auxiliary quantity given
in \eqref{gamma}. Here, $\bar{u}^{a}$ is the matter velocity and
$\bar{x}^{a}$, $\bar{y}^{a}$ and $\bar{z}^{a}$ are spacelike
eigenvectors of the pulled-back material metric $\bar{k}^{a}_{b}$.
From our developments in section 2, it is immediate to see that the
pressure tensor has the same eigenvectors as $\bar{k}_{ab}$ and can
be written, for the space-time under consideration as
$\bar{P}_{ab}=\bar{p}_1\bar{x}_{a}\bar{x}_{b}+\bar{p}_2(\bar{y}_{a}\bar{y}_{b}+\bar{z}_{a}\bar{z}_{b})$.
Therefore, \eqref{2.2} yields
\begin{equation} \label{20} \bar{T}_{ab}=\bar{\rho}
\bar{u}_{a}\bar{u}_{b}+\bar{p}_{1}\bar{x}_{a}\bar{x}_{b}+\bar{p}_{2}(\bar{y}_{a}\bar{y}_{b}+\bar{z}_{a}\bar{z}_{b}),
\end{equation} where $\bar{\rho}$ is the energy density, $\bar{p}_{1}$,
the radial pressure and $\bar{p}_{2}$, the tangential pressure.

The results in \cite{M1} can be easily recovered by setting $f(y)=1$
above.

Now, much clarity is gained by making use of the comoving
coordinates adapted to $\vec u$, the timelike eigenvector of the
energy-momentum tensor, which were introduced  in the above section.
The form of the metric is given by \be ds^2= -\bar a(r,t) dt^2 +
\bar b(r,t) dr^2+ \bar Y^2(r,t)\left( d \theta^2 + \sin^2 \theta d
\phi^2\right), \label{metrica2b} \ee  $\vec u$, being then
\be\label{u1} u^a = \left(\frac{1}{\sqrt{\bar a}},0,0,0\right),
\qquad u_a = \left(-\sqrt{\bar a},0,0,0\right),\ee hence, we have
for the material space $(M,\bar \gamma)$ that coordinates $y^A =
(y,\tilde\theta,\tilde\phi )$ exist with $y=y(r)$,
$\tilde{\theta}=\theta$ and $\tilde{\phi}=\phi$, as follows from the
condition  $y^A_{a}u^a =0$ and the requirement that $\psi_*(\vec
\xi_A) = \vec\eta_A$ are KVs of the metric $\bar\gamma$.

Further, and since the line element of the material space is
$$ d\bar\sigma^2 = f^2(y)\left[dy^2 + y^2\left(d\theta^2 +\sin^2\theta d\phi^2\right)\right],$$
with $y=y(r)$, no generality is lost by setting $y=r$, as this amounts to a redefinition of the
$r$ coordinate in spacetime, and leaves unchanged the form of the metric \eqref{metrica2b} as well
as that of the velocity field of the matter \eqref{u1}. We shall do that in the sequel.

Thus, the pulled-back material metric $\bar k$ \eqref{22} is
\begin{equation*}
\begin{split}
\bar{k}^{a}_{b}
&=\bar{g}^{ac}\bar{k}_{cb}=\bar{g}^{ac}\bar{\gamma}_{CB}y^{C}_{c}y^{B}_{b}=f^{2}(y)\bar{g}^{ac}\gamma_{CB}
y^{C}_{c}y^{B}_{b}=f^{2}(y)\bar{g}^{ac}{k}_{cb}\\
&=f^{2}(y)\bar{g}^{ac}[
y'^{2}\delta^{1}_{b}\delta^{1}_{c}+y^{2}\delta^{2}_{b}\delta^{2}_{c}+y^{2}sin^{2}\theta
\delta^{3}_{b}\delta^{3}_{c}],
\end{split}
\end{equation*}
where  a prime indicates a derivative with
respect to $r$, which upon setting  $y=r$ as discussed above it simplifies further to:
\begin{equation}\bar{k}^{a}_{b} = \left(\begin{array}{cccc}0 &0&0&0\\
0&f^{2}(r)(1/\bar{b})&0&0\\0&0&f^{2}(r)r^{2}/Y^{2}&0\\0&0&0&f^{2}(r)r^{2}/Y^{2}\end{array}
\right).\label{pulledbk}\end{equation}

The operator
$\bar{K}^{a}_{b}=\bar{g}^{ac}\bar{k}_{cb}-\bar{u}^{a}\bar{u}_{b}$,
introduced in section 2 and used to measure the state of strain of
the material has one eigenvalue equal to 1 (corresponding to
the eigenvector $\vec u$), while the other eigenvalues are
\begin{equation}\begin{split}\label{etas}\bar{s}&=\,f^{2}(y)\,\frac{y^{2}}{\bar{Y}^{2}} =
f^{2}(r)\,\frac{r^{2}}{\bar{Y}^{2}}\\ \bar{\eta}&=f^{2}(y)\,
\frac{y'^{2}}{\bar{b}}=\frac{f^{2}(r)}{\bar{b}},\end{split}\end{equation} and $\bar{s}$ has algebraic multiplicity two.\\

The three invariants $I_1, I_2, I_3$ of $\bar{K}$ introduced in
\eqref{I1.0} have the following expressions
\begin{equation}\begin{split}\label{I1}\bar{I_{1}}&=\frac{1}{2}\left(\text{Tr}\bar{K}-4\right) =
\frac{1}{2}\left(\bar\eta + 2\bar s -3\right)\\
\bar{I_{2}}&=\frac{1}{4}\left[\text{Tr}\bar{K}^{2}-\left(\text{Tr}\bar{K}\right)^{2}\right]+3=
 -\frac12 \left(\bar s^2 +2\bar \eta\bar s +\bar\eta +2\bar s\right) -3\\
\bar{I_{3}}&=\frac{1}{2}\left(\text{det}\bar{K}-1\right) =
\frac12\left( \bar\eta\bar s^2 -1\right)\end{split}\end{equation}

In \cite{M1}, the energy-momentum tensor was calculated from these
invariants for a flat material metric. A similar calculation shows
that, for the non-flat material metrics under consideration, the
same expression holds so that
\begin{equation} \label{25-Tab}
\bar{{T}}^{a}_{b}=\bar{\rho}\,\delta^{a}_{b}-\frac{\partial\bar{\rho}}{\partial\bar{I}_{3}}\,
\text{det}\bar{K}\,\bar{h}^{a}_{b}+ \left(\text{Tr}
\bar{K}\,\frac{\partial\bar{\rho}}{\partial\bar{I}_{2}}-\frac{\partial\bar{\rho}}{\partial\bar{I}_{1}}\right)\bar{k}^{a}_{b}-
\frac{\partial\bar{\rho}}{\partial\bar{I}_{2}}\,\bar{k}^{a}_{c}\,\bar{k}^{c}_{b}.
\end{equation}
Therefore, the nonzero components are
\begin{equation*}
\bar{{T}}^{0}_{0}=\,\bar{\rho},\end{equation*}
\begin{equation*}
\bar{{T}}^{1}_{1}=\,\bar{\rho}-\frac{y'^{2}}{\bar{b}}\,\overline{\sum},\end{equation*}
\begin{equation}
\bar{{T}}^{2}_{2}=\bar{T}^{3}_{3}=\,\bar{\rho}-\frac{y^{2}}{\bar{Y}^{2}}\,\left[\,\overline{\sum}\,+\,
\left(\frac{\partial\bar{\rho}}{\partial\bar{I}_{2}}\,-\,f^{2}(y)\,\frac{y^{2}}{\bar{Y}^{2}}\,\frac{\partial\bar{\rho}}{\partial\bar{I}_{3}}\right)
\left(f^{4}(y)\,\frac{y^{2}}{\bar{Y}^{2}}-f^{4}(y)\,\frac{y'^{2}}{\bar{b}}\right)\right],
\label{T1}\end{equation} where \be
\overline{\sum}=f^{2}(y)\left[\frac{\partial\bar{\rho}}{\partial\bar{I}_{1}}-
\frac{\partial\bar{\rho}}{\partial\bar{I}_{2}}\left(\,1\,+\,2\,f^{2}(y)\,\frac{y^{2}}{\bar{Y}^{2}}\right)\,+\,
\frac{\partial\bar{\rho}}{\partial\bar{I}_{3}}\,f^{4}(y)\,\frac{y^{4}}{\bar{Y}^{4}}\right].
\ee The rest frame energy per unit volume{\footnote{In \cite{M1} the
quantities $\bar{\rho}$ and $\bar{\epsilon}$ are $\epsilon$ and
$\rho$, respectively. }}, $\bar{\rho}$, is defined by \be
\bar{\rho}=\bar{\epsilon}\bar{v}=\,\epsilon_{0}\,\bar{s}\,\sqrt{\bar{\eta}}\,\bar{v}(\bar{s},\bar{\eta}),
\label{rfe}\ee where, as discussed in section 2,
$\bar{v}=\bar{v}(\bar{I}_{1},\bar{I}_{2},\bar{I}_{3})=\bar{v}(\bar{s},\bar{\eta})$
represents the constitutive equation, $\epsilon_{0}$, the density of
the relaxed material (density w.r.t the pulled-back material metric
$\bar k$) and \be \label{drm} \bar{\epsilon}\,=\,
\epsilon_{0}\,\sqrt{\text{det}\bar{K}}=\,\epsilon_{0}\,\bar{s}\sqrt{\bar{\eta}},
\ee the density calculated in the rest frame (that is, w.r.t. $h$).

Then, using \eqref{I1}, one can prove the following relations:
\begin{align} &\frac{\partial
\bar{\rho}}{\partial\bar{\eta}}=\frac{1}{2\,f^{2}}\,\overline{\sum},\\
&\frac{\partial\bar{\rho}}{\partial\bar{s}}=\frac{1}{f^{2}}\,\overline{\sum}\,+\,
\left(\,f^{2}\,\frac{\partial\bar{\rho}}{\partial\bar{I}_{2}}\,-
\,f^{4}\,\frac{y^{2}}{\bar{Y}^{2}}\,\frac{\partial\bar{\rho}}{\partial\bar{I}_{3}}\right)\,
\left(\frac{y^{2}}{\bar{Y}^{2}}- \frac{y'^{2}}{\bar{b}}\right).
\end{align}
Alternatively, one can express the components of the energy-momentum
tensor in terms of the eigenvalues $\bar{s}$ and $\bar{\eta}$ by
substituting the last results in \eqref{T1}:
\begin{equation}\begin{split}
\label{T2}
&\bar{{T}}^{0}_{0}= \bar{\epsilon}\bar v,\\
&\bar{{T}}^{1}_{1}=-\bar{\epsilon}\,
2\,\bar{\eta}\,\frac{\partial\bar{v}}{\partial\bar{\eta}},\\
&\bar{{T}}^{2}_{2}=-\bar{\epsilon}\,\bar{s}\,\frac{\partial\bar{v}}{\partial\bar{s}}.
\end{split}\end{equation}

The Einstein field equations $\bar{G}^{a}_{b}=8\pi
\bar{{T}}^{a}_{b}$ can be written as follows:

$\bar{G}^{0}_{0}=8\pi \bar{{T}}^{0}_{0}$: \be
-\frac{\dot{\bar{Y}}}{\bar{Y}^{2}\bar{a}}-\frac{\dot{\bar{Y}}}{\bar{Y}}\frac{\dot{\bar{b}}}{\bar{a}\bar{b}}+
\frac{2\bar{Y}''}{\bar{Y}\bar{b}}+\frac{\bar{Y}'^{2}}{\bar{Y}^{2}\bar{b}}-
\frac{\bar{Y}'}{\bar{Y}}\frac{\bar{b}'}{\bar{b}^{2}}-\frac{1}{\bar{Y}^{2}}=
\bar{\epsilon}\bar v\,8\pi\ee

$\bar{G}^{1}_{0}=8\pi \bar{{T}}^{1}_{0}$: \be
2\dot{\bar{Y}}'-\frac{\bar{a}'}{\bar{a}}\dot{\bar{Y}}-\frac{\dot{\bar{b}}}{\bar{b}}\bar{Y}'=0,
\ee

$\bar{G}^{1}_{1}=8\pi \bar{{T}}^{1}_{1}$: \be
-\frac{\dot{\bar{Y}}^{2}}{\bar{Y}^{2}\bar{a}}+\frac{\dot{\bar{Y}}}{\bar{Y}}\frac{\dot{\bar{a}}}{\bar{a}^{2}}+
\frac{\bar{Y}'}{\bar{Y}}\frac{\bar{a}'}{\bar{a}\bar{b}}+\frac{\bar{Y}'^{2}}{\bar{Y}^{2}\bar{b}}-
\frac{2\ddot{\bar{Y}}}{\bar{Y}\bar{a}}-\frac{1}{\bar{Y}^{2}}=-\bar{\epsilon}\,
2\,\bar{\eta}\,\frac{\partial\bar{v}}{\partial\bar{\eta}}\,8\pi, \ee

$\bar{G}^{2}_{2}=8\pi \bar{{T}}^{2}_{2}$: \be\begin{split}
&\frac{1}{2}\frac{\dot{\bar{Y}}\dot{\bar{a}}}{\bar{Y}\bar{a}^{2}}-
\frac{1}{2}\frac{\dot{\bar{Y}}\dot{\bar{b}}}{\bar{Y}\bar{a}\bar{b}}-\frac{1}{4}\frac{\bar{a}'^{2}}{\bar{a}^{2}\bar{b}}+
\frac{1}{2}\frac{\bar{Y}'\bar{a}'}{\bar{Y}\bar{a}\bar{b}}-\frac{1}{4}\frac{\bar{a}'\bar{b}'}{\bar{a}\bar{b}^{2}}+
\frac{\bar{Y}''}{\bar{Y}\bar{b}}-\frac{1}{2}\frac{\bar{Y}'\bar{b}'}{\bar{Y}\bar{b}^{2}}+\\
&\frac{1}{2}\frac{\bar{a}''}{\bar{a}\bar{b}}-\frac{1}{2}\frac{\ddot{\bar{b}}}{\bar{a}\bar{b}}+
\frac{1}{4}\frac{\dot{\bar{a}}\dot{\bar{b}}}{\bar{a}^{2}\bar{b}}+\frac{1}{4}\frac{\dot{\bar{b}}^{2}}{\bar{a}\bar{b}^{2}}-
\frac{\ddot{\bar{Y}}}{\bar{Y}\bar{a}}=\\
&-\bar{\epsilon}\,\bar{s}\,\frac{\partial\bar{v}}{\partial\bar{s}}\,8\pi.
\end{split}\ee

It is interesting to express the contracted Bianchi identities for $
\bar{{T}}^a_{\;b}$ in terms of $\bar v$ and its derivatives w.r.t
the quantities $\bar\eta$ and $\bar s$. Thus, from $\bar T^{a}_{\;
b;a}=0$ one has: \be\label{bianchi1}\bar T^{a}_{\;
b,a}+\partial_{n}(\ln \sqrt{-\bar g})\bar
T^{n}_{b}-\bar\Gamma^{n}_{ba}\bar T^{a}_{n}=0 \ee and specifying
this equation to $b=0, 1$ one gets respectively (for non-stationary
solutions): \be\label{bianchi2}
\partial_{t}(\bar\epsilon\bar v)+\frac{\dot{\bar b}}{\bar b}\left(\frac{1}{2}\bar\epsilon \bar v+
\bar\epsilon\bar\eta\frac{\partial \bar v}{\partial\bar\eta}\right)
+\frac{\dot{\bar Y}}{\bar Y}(2\epsilon \bar v+2\bar \epsilon \bar
s\frac{\partial \bar v}{\partial \bar s})=0,\ee  \be\label{bianchi3}
-2\left(\bar\epsilon \bar\eta \frac{\dep \bar
v}{\dep\bar\eta}\right)_{,r}-\frac12 \bar\epsilon\bar v\frac{\bar
a'}{\bar a}- \bar\epsilon\bar\eta \frac{\partial\bar
v}{\partial\bar\eta}\left(\frac{\bar a'}{\bar a}+4\frac{ \bar
Y'}{\bar Y}\right) + 2\bar\epsilon \bar s\frac{\partial \bar
v}{\partial \bar s}\frac{\bar Y'}{\bar Y}=0.\ee The remaining
equations for $b=2,3$ which can be obtained from \eqref{bianchi1}
are identically satisfied.

Equation \eqref{bianchi2} for non-stationary solutions, implies
readily \be\label{bianchi4} \bar\epsilon=\frac{1}{\sqrt{\bar b}\bar
Y^{2}}\epsilon_{0}(r),\ee which can then be substituted into
(\ref{bianchi3}) to get a slightly simplified equation.

From this point onwards, we shall drop the bars, as no confusion may
arise with the results in \cite{M1}.

\section{Shearfree solutions. Examples}
In this section we shall consider in detail the case of spacetimes
with a material content that may be represented by some elastic material such that
the velocity of the matter is shearfree, in which case coordinates exist such that the
metric can be written in the form \eqref{metrica3}. For this case,
the interpretation as a viscous fluid with kinematical viscosity is
not possible, and therefore the anisotropy in the pressures must be
a consequence of the elastic properties of the material.  The study of
solutions with non-vanishing shear tensor and their possible interpretations
as viscous fluids, will be carried out elsewhere
as this would render the present paper too lengthy.

We will study separately the cases of static and non-static solutions,
presenting  examples of each instance which are regular at the origin, posses an equation of state
and satisfy the dominant energy condition (at least in some open submanifold of
the spacetime).

Consider the metric \eqref{metrica3} which we rewrite here for
convenience: \be\label{shearfree0} ds^2= -a(r,t) dt^2 +
Y^2(r,t)\left( dr^2+  d \theta^2 + \sin^2 \theta d \phi^2\right)\ee
From the field equations it follows that $G^t_{\; r} =0$ which in
turn implies that \be\label{shearfree1} a = L(t) \frac{\dot Y}{Y}\ee
whenever $\dot Y \neq 0$, $L(t)$ being a function of time.

If $\dot Y = 0$ then $G^t_{\; r} = 0$ is identically satisfied, and
from \eqref{etas} follows that $s$ and $\eta$, and therefore
$v(\eta,s)$ are functions of $r$ alone, then \eqref{bianchi2}
implies that $\epsilon = \epsilon (r)$; further from the field
equation $G^r_{\;r} = -8\pi \epsilon 2 \eta \frac{\partial
v}{\partial \eta}$ it follows that  $G^r_{\;r}$ can only depend on
$r$ as well, which in turn implies that $a(t,r) = a_0(t)a_1(r)$, the
solution being then static, as a trivial redefinition of the
coordinate $t$ coordinate shows.

It is interesting now to see that in the shear-free case, if one
sets either $\eta$ or $s$ equal to 1, so that matter is strained in
tangential directions (but not in the  radial direction $\eta =1$),
or it is strained only in the radial direction ($s=1$), from the
definition of these quantities it follows that $Y=Y(r)$, and
according to the statements in the above paragraph, it follows that
the solution must be static, and therefore the results in
\cite{Park} apply. Thus, we have proven that: \emph{if the velocity
field of the matter is shear-free and the matter is stressed either
in the radial direction only or in the tangential directions only,
the spacetime is necessarily static.}

\subsection{Static shearfree solutions}
In the static, shear-free case (metric \eqref{shearfree0} with
no dependence on $t$), the field equations yield
\be\label{shearfreestatic2}  \epsilon v 8\pi= 2\frac{Y''}{Y^3} -
\frac{Y'^2}{Y^4}-\frac{1}{Y^2},\ee \be\label{shearfreestatic3}
-2\epsilon \eta\frac{\dep v}{\dep\eta} 8\pi=
 \frac{a'}{a} \frac{Y'}{Y^3} + \frac{Y'^2}{Y^4}-\frac{1}{Y^2},\ee
\be\label{shearfreestatic4} -\epsilon s\frac{\dep v}{\dep s}8\pi =
 \frac{Y''}{Y^3} + \frac12\frac{a''}{aY^2}-\frac14\frac{a'^2}{a^2Y^2}-\frac{Y'^2}{Y^4},\ee solving
\eqref{shearfreestatic2} for $\epsilon$ and substituting it in
\eqref{shearfreestatic3} and \eqref{shearfreestatic4} one gets two
equations which depend only on $r$ and elementary considerations
show that for given $a(r)$ and $Y(r)$, functions $y(r)$,  $f(y)$ and
$v$ can be found so that the two equations are satisfied. It remains
to be seen, though, that the DEC are satisfied and therefore the
solution is physically acceptable.

The following simple example shows that solutions with these characteristics
do indeed exist.

\textbf{Example 1}

Consider the line element \be\label{example1.0} ds^2= -Y^{-2}(r)
dt^2 + Y^2(r)\left( dr^2+  d \theta^2 + \sin^2 \theta d
\phi^2\right).\ee A direct calculation yields \be\label{example1.1}
8\pi \epsilon v = 2\frac{Y''}{Y^3} - \frac{Y'^2}{Y^4}-\frac{1}{Y^2},
\quad   -16\pi\epsilon \eta\frac{\dep v}{\dep\eta} = -
\frac{Y'^2}{Y^4}-\frac{1}{Y^2}, \quad -8\pi\epsilon s\frac{\dep
v}{\dep s} = \frac{Y'^{2}}{Y^{4}}.\ee

The dominant energy condition \eqref{2.3} implies: \be\label{DEC1}
8\pi \,\rho= 2\frac{Y''}{Y^3} - \frac{Y'^2}{Y^4}-\frac{1}{Y^2}\geq 0,\ee
\be\label{DEC2} 8\pi \,(\rho -p_1) = 2\frac{Y''}{Y^3}\geq 0, \ee
\be\label{DEC3} 8\pi \,(\rho + p_1) = 2\left(\frac{Y''}{Y^3} -
\frac{Y'^2}{Y^4}-\frac{1}{Y^2}\right)\geq 0\ee \be\label{DEC4}
8\pi \,(\rho-p_2) =2\frac{Y''}{Y^3} - 2\frac{Y'^2}{Y^4}-\frac{1}{Y^2}\geq 0,
\ee \be\label{DEC5}8\pi \,(\rho+p_2) =2\frac{Y''}{Y^3}-\frac{1}{Y^2}\geq
0,\ee where we put $\rho = \epsilon v$, $p_1 = -2\epsilon
\eta\frac{\dep v}{\dep\eta}$ and $p_2 =-\epsilon s\frac{\dep v}{\dep
s}$.

Now, it is immediate to see that the above conditions are all
satisfied if and only if \eqref{DEC3} is, which in turn can be
written as \be\label{DEC3.1} \frac1{Y^2}\left( \frac{Y''}{Y}
-\frac{Y'^2}{Y^2} -1\right)\geq 0 \qquad \Leftrightarrow \qquad (\ln
Y)'' -1 \geq 0,\ee which is equivalent to \be\label{DEC3.2} Y
=\exp\left(r^2/2\right) f^2(r)\qquad \mathrm{{such \;that}}\qquad
(\ln f)''\geq 0.\ee

Take, for instance, \be\label{example1.2} Y = \exp(5/2 r^2),\ee one
then has \be\nonumber \rho = \epsilon v = \frac1{8\pi} e^{-5r^2}(25 r^2 +9),\qquad\qquad\qquad\\
\qquad p_1 =-2\epsilon \eta\frac{\dep v}{\dep\eta}= -
\frac1{8\pi}e^{-5r^2}(25 r^2 + 1), \qquad p_2 =-\epsilon s\frac{\dep
v}{\dep s} =\frac1{8\pi} 25r^2 e^{-5r^2}\label{example1.3} \ee which
is obviously well behaved: satisfies the dominant energy condition
and is non-singular at the origin. Notice that the radial pressure
is negative (compressed material) and the tangential pressures are
zero at the centre, as one would expect.

The field equations in this case read:
\be\label{EFE1.example1} \epsilon v =\frac1{8\pi} e^{-5r^2}(25r^2 + 9)\ee
\be\label{EFE2.example1} -2\epsilon\eta\frac{\dep v}{\dep\eta} =-\frac1{8\pi} e^{-5r^2}(25r^2 + 1)\ee
\be\label{EFE3.example1} -\epsilon s\frac{\dep v}{\dep s} =\frac1{8\pi} e^{-5r^2}25r^2 \ee
and one has that
\be\label{etas.example1} \eta = f^2(r) e^{-5r^2}, \qquad s = r^2 f^2(r) e^{-5r^2}.\ee

Now, dividing  \eqref{EFE2.example1}   and \eqref{EFE3.example1}
through by \eqref{EFE1.example1}, and setting $E\equiv \ln\eta, \,
\Sigma \equiv \ln s$, one gets \be\label{EFE4} \frac{\dep \ln
v}{\dep E}= \frac12 \frac{25r^2+1}{25r^2+9}, \qquad \frac{\dep \ln
v}{\dep \Sigma}= - \frac{25r^2}{25r^2+9}.\ee

From the expressions for $\eta$ and $s$ one has that
\be\label{ESigma} E = 2 \ln f(r) -5r^2, \qquad \Sigma =E + 2\ln
r,\ee hence one can express $r$ as a function of $E$, and
$\Sigma$ as a function of $E$ as well, thus $$ \frac{\dep
\ln v}{\dep E}= \frac{\dep \Sigma}{\dep E}\frac{\dep \ln v}{\dep
\Sigma}$$ from where it follows that $$ \frac{\dep r}{\dep E} =
-\frac34 r-\frac1{100r}, \quad \mathrm{or \, else} \quad E =
-\frac23 \ln(75r^2+1),$$ that is \be\label{rE} r =
\sqrt{\frac1{75}\left(e^{-\frac32 E} -1\right)}.\ee Plugging the
expression of $E$ in terms of $r$ into \eqref{ESigma} one gets that
\be\label{f.example1} f(r) = \frac{e^{\frac52
r^2}}{(75r^2+1)^{\frac13}},\ee whence expressions for $\eta$, $s$
and $\epsilon = \epsilon_0 s\sqrt{\eta}$ can be easily  derived.

Next, from \eqref{rE} and the first equation in \eqref{EFE4}, one
can easily find an expression for the equation of state, namely:
\be\label{v.example1} v = F(\Sigma)\left(\frac{e^{\frac{3}{2}
E}}{\left(e^{-\frac{3}{2} E} +
26\right)^{12}}\right)^{\frac1{39}},\ee where $F(\Sigma)$ must
satisfy
$$ \frac{\dep \ln F}{\dep\Sigma} = -  \frac{25r^2}{25r^2+9},$$
where $r$ in the right hand side of the equation has to be expressed in terms of $\Sigma$.
From the second equation in \eqref{ESigma}
it follows that $r$ must be the only real solution of $$ r^3 -75 e^{\frac32 \Sigma} r^2 - e^{\frac32 \Sigma}=0,$$
which has a rather complicated form. In any case, one gets
$$ F(\Sigma(r)) = \left( 25r^2 +9 \right)^{-\frac{12}{39}} \left( 75r^2 +1 \right)^{-\frac{1}{39}},$$
and thus we have proven that a solution exists, which is regular at the origin $r=0$,
satisfies the dominant energy condition and
possesses an equation of state which can be given in a closed form.

\subsection{Non-static shearfree solutions}
Assume now that $\dot Y \neq 0$, so that $a(t,r)$ takes the form
\eqref{shearfree1}, substituting this into \eqref{shearfree0} and
redefining the coordinate $t$ so as to absorb the arbitrary function
$L(t)$ one has \be\label{shearfree2} ds^2 = - \frac{\dot Y}{Y} dt^2
+ Y^2(r,t)\left(dr^2 + d \theta^2 + \sin^2 \theta d \phi^2\right)\ee

From $G^t_{\;r}=0$ it follows now that $Y(r,t) = A(t)B(r)$, which
substituted above, yields, after a trivial redefinition of the
coordinate $t$: \be\label{shearfree3} ds^2 = - dt^2 +
A^2(t)B^2(r)\left(dr^2 + d \theta^2 + \sin^2 \theta d
\phi^2\right)\ee

A direct computation of the EFEs for the above metric give gives
\be\label{efe00} 8\pi \epsilon
v=-\frac{B'^{2}-2B''B+3\dot{A}^{2}B^{4}+B^{2}}{A^{2}B^{4}},\ee
\be\label{efe11} -16\pi\epsilon\eta\frac{\partial
v}{\partial\eta}=-\frac{2AB^{4}\ddot{A}-B'^{2}+\dot{A}^{2}B^{4}+B^{2}}{A^{2}B^{4}},\ee
\be\label{efe22} -8\pi\epsilon s\frac{\partial v}{\partial
s}=-\frac{2AB^{4}\ddot{A}-B''B+\dot{A}^{2}B^{4}+B'^{2}}{A^{2}B^{4}}.\ee

where the energy density, radial and tangential pressures are $$\rho =\epsilon
v, \qquad p_1 = -2\epsilon\eta\frac{\partial
v}{\partial\eta}, \qquad p_2 = -\epsilon s\frac{\partial v}{\partial
s}.$$

On the other hand, the dominant energy condition,
$\rho \geq 0, \rho \pm p_1 \geq 0$ and $\rho \pm p_2 \geq 0$ implies

\be\label{dec00}
- (B'^{2}-2B''B+3\dot{A}^{2}B^{4}+B^{2})\geq
0,\ee
\be\label{dec01minus}
-B'^{2}+B''B-\dot{A}^{2}B^{4}+AB^{4}\ddot{A} \geq
0,\ee
\be\label{dec01plus}
-(AB^{4}\ddot{A}-B B''+2\dot{A}^{2}B^{4}+B^2)\geq
0,\ee
\be\label{dec02minus}
B B''-2\dot{A}^{2}B^{4}-B^2+2AB^{4}\ddot{A}\geq
0,\ee
\be\label{dec02plus}
- (2B'^{2}-3B''B+4\dot{A}^{2}B^{4}+B^{2}+2AB^{4}\ddot{A}) \geq
0.\ee

We next address the question of the existence of an equation of state $v = v(\eta, s)$ where in this case, $\eta$ and $s$ are given by (see \eqref{etas})
\be\label{etas2} \eta = \frac{f^2(r)}{A^2(t)B^2(r)}, \qquad s =  \frac{r^2 f^2(r)}{A^2(t)B^2(r)}. \ee

Dividing \eqref{efe11} and \eqref{efe22} by \eqref{efe00}, and defining as before $E = \ln \eta$, $\Sigma =\ln s$, we get
\be\label{VE} \frac{\dep \ln v}{\dep E} =-\frac12 \frac{2AB^{4}\ddot{A}-B'^{2}+\dot{A}^{2}B^{4}+B^{2}}{B'^{2}-2B''B+3\dot{A}^{2}B^{4}+B^{2}} \equiv V_E, \ee

\be\label{VS} \frac{\dep \ln v}{\dep \Sigma} =- \frac{2AB^{4}\ddot{A}-B''B+\dot{A}^{2}B^{4}+B'^{2}}{B'^{2}-2B''B+3\dot{A}^{2}B^{4}+B^{2}} \equiv V_S. \ee

In order for an equation of state $v=v(\eta,s)$ (or equivalently $v= v(E,\Sigma)$) to exist, it must be that
\be\label{existence0} \frac{\dep^2 \ln v}{\dep \Sigma \dep E} = \frac{\dep^2 \ln v}{\dep E \dep \Sigma} \quad \Leftrightarrow \quad \frac{\dep V_E}{\dep\Sigma} = \frac{\dep V_\Sigma}{\dep E}.\ee

Notice that \be\label{chainrule} \dep_E = \frac{\dep t}{\dep E}\,\dep_t + \frac{\dep r}{\dep E}\,\dep_r, \qquad \dep_\Sigma = \frac{\dep t}{\dep\Sigma}\,\dep_t + \frac{\dep r}{\dep\Sigma}\,\dep_r.                \ee

Now, from the expression \eqref{etas2} and the corresponding one for $E$ and $\Sigma$, it follows that
\be\label{rt} r = e^{\frac12(\Sigma -E)}, \qquad A(t)= \frac{f}{B} e^{-\frac12 E}; \ee
and differentiating them with respect to $E$ and $\Sigma$ and applying the chain rule, we get
\be\label{derivatives1} \frac{\dep r}{\dep E} = -\frac12 r, \qquad \frac{\dep r}{\dep \Sigma} = \frac12 r, \ee

\be\label{derivatives2} \frac{\dep t}{\dep E} = -\frac1{2\dot A} \left[ r\left( \frac{f}{B}\right)' +  \frac{f}{B}\right] \frac{A B}{f}, \qquad \frac{\dep t}{\dep \Sigma} = \frac1{2\dot A} r\left( \frac{f}{B}\right)'  \frac{A B}{f}.\ee

Substituting the above expressions into \eqref{chainrule}, equation \eqref{existence0} reads, after some manipulations
\be\label{existence1} \frac{A}{\dot A}\, r \left( \frac{f'}{f} -  \frac{B'}{B} \right)\dep_t (V_E +V_S) + r \, \dep_r (V_E + V_S) + \frac{A}{\dot A}\, \dep_t V_S = 0.\ee

Solving for $\displaystyle{\frac{f'}{f}}$ we get, after some algebra,
\be\label{existence2} 3rB^2 \frac{f'}{f} = \frac{-K\left[ (S+ 3 B^4 \dot A^2) A_{ttt} - 6 B^4 \dot A \ddot A^2 + 3 B^4 A^{-1}\dot A^3  \right] + M A^{-1} \dot A \ddot A }{\left[ (S+ 3 B^4 \dot A^2) A_{ttt} - 6 B^4 \dot A \ddot A^2 + 3 B^4 A^{-1}\dot A^3  \right] + S A^{-1} \dot A \ddot A },\ee

where, $K, S$ and $M$ are functions of $r$ alone given by:
$$ K = 2B -3rB', \quad S = B'^2 -2B B'' + B^2, $$
$$ M = -6r B^2 B_{rrr} + 2B(9r B' + B)B'' + (2 B B' -9r B'^2 - 3r B^2) B' -4B^3.$$

Since the left hand side of \eqref{existence2} depends only on $r$ this implies that the time derivative of the right hand side must vanish. A careful but otherwise trivial analysis, reveals that there are only three possibilities, assuming the metric is n
 on-static, namely
\begin{enumerate}
  \item  $ M+ KS = 0$, in which case $B(r)$ is determined by the resulting ordinary third order differential equation. $A(t)$ is in principle arbitrary, and $f(r)$ is fixed by \eqref{existence2} once the solution for $B(r)$ to the equation $M+KS=0$ is give
 n. We have not been able to find an integral for $B(r)$ in closed form, but in this case one has for $f(r)$:
  $$  \frac{f'}{f} = \frac{B'}{B^2}-\frac{2}{3rB}.$$
  \item $\ddot A =0$, which in turn implies $A = t$ without loss of generality, since the two constants of integration may be absorbed by suitable redefinitions of $t$ and $B$. In this case, $B(r)$ is free, constrained only by the requirements imposed by t
 he DEC, and once it is chosen, $f(r)$ is determined through \eqref{existence2}, which implies as in the previous case  $$  \frac{f'}{f} = \frac{B'}{B^2}-\frac{2}{3rB}.$$
  \item  In this case, both $A(t)$ and $B(r)$ are determined as the solutions of the following two third order differential equations:
  $$ -2BB'' + B'^2 + B^2-k B^4 =0,  \qquad k = \mathrm{constant},$$
  $$ A A_{ttt} \left(\frac{k}{3} + \dot A \ddot A\right) - 2A \dot A \ddot A^2 + \dot A^3 -q \dot A\ddot A = 0, \qquad k, q = \mathrm{constant}.$$
  Since $A(t)$ and $B(r)$ are fixed, so is $f(r)$, and \eqref{existence2} implies in this case
$$  3r B^2 \frac{f'}{f} = \frac{M + (k-3q) K B^4}{3q B^4}.$$
As in the first case above, we have not been able to find integrals for $A(t)$ or $B(r)$ in closed form.
\end{enumerate}

\textbf{Example 2}

Let us next investigate in some detail the second case above, that is $A =t$. Substituting this into the EFEs we get
\be\label{efeexample2} \rho =\frac1{8\pi\, t^2}\left( \frac{2B''}{B^3} - \frac{B'^2}{B^4} - \frac{1}{B^2} - 3  \right), \quad p_1 = \frac1{8\pi\, t^2} \left( \frac{B'^2}{B^4} - \frac{1}{B^2} - 1  \right), \quad \quad p_2 = \frac1{8\pi\, t^2} \left( \frac{B''}{B^3}- \frac{B'^2}{B^4} -  1  \right). \ee

On the other hand, the DEC \eqref{dec00}-\eqref{dec02plus} are all satisfied if and only if the following three inequalities hold:
\be \label{decexample2a} 3BB'' -2 B'^2 -B^2 -4B^4\geq 0 , \quad B B'' - B'^2 - B^4 \geq 0 , \quad B B'' - B^2 - 2 B^4 \geq 0 \quad  \ee
which, upon setting $B \equiv e^b$  are equivalent to
\be \label{decexample2b} 3b'' + b'^2 -1 - 4 e^{2b}  \geq 0,  \qquad    b'' -e^{2b}  \geq 0,  \qquad  b'' + b'^2 -1 -2 e^{2b}  \geq 0. \ee

It is easy to see that these conditions can be satisfied, at least for certain fanges of the radial coordinate $r\in [0, R)$, for suitably chosen functions $b(r)$, such as

\be\label{b1} b(r) = \frac32 \ln \left[\frac23 +  \sinh^2 \left(\frac{r-r_0}{c}\right)\right] - \ln c \quad \Leftrightarrow \quad B(r) = \frac{\sqrt{3}}{9c} \ln \left[-1 + 3 \cosh^2 \left(\frac{r-r_0}{c}\right)\right]^{\frac32}\ee

The form of $f(r)$ can be given explicitly up to a quadrature.

While we do not claim that it has any particular significance, it provides a relatively simple instance of solution
with the desired properties.

\textbf{Example 3}

Another simple example with similar  characteristics is provided by the following choice of $B(r)$:

\be\label{b2} b(r) = \frac32 \ln\left(\frac23 +r^2\right) \quad \Leftrightarrow \quad B(r) = \frac{\sqrt{3}}{9} \left(2 + 3 r^2\right)^\frac32, \ee
 in which case $f(r)$ can be integrated out yielding:

$$ f(r) =  \exp\left\{ \frac{      -15-9 {r}^{2} + \frac32  (2+ 3r^2) \sqrt{4 + 6 r^2}    \tanh^{-1}  \left(  \frac{2 }{ \sqrt{4 + 6 r^2} }  \right)  }{\sqrt {6+9\,{r}^{2}} \left( 2+3\,{r}^{2} \right) }  \right\}$$

Similar remarks to the ones in the previous case regarding its physical significance, apply also here.

\vskip 18pt

\section{The elasticity difference tensor for non-static solutions}

Here we obtain the \edt, defined in \cite{KS1}, for non-static
spherically symmetric spacetimes and analyze this tensor following
the procedure developed in \cite{VB}, where the static, spherically
spacetime case was presented as an example.

This third order tensor, symmetric on the two covariant indices, is
completely flow-line orthogonal and is related with the (pulled back)
material metric according to \be
S^{a}_{\hspace{0.15cm}bc}=\frac{1}{2}k^{-am}(D_{b}k_{mc}+D_{c}k_{mb}-D_{m}k_{bc}).\ee
Here $k^{-am}$ is such that $k^{-am}k_{mb}=h^{a}_{b}$ and $D$
represents the spatially projected connection obtained from the
spacetime connection $\nabla$ associated with $g$ by \be
D_{a}t^{b...}_{c...}=h^{d}_{a}h^{b}_{e}...h^{f}_{c}...\nabla_{d}t^{e...}_{f...}\,
, \ee with the property $D_{a}h_{bc}=0$.


The non zero components of $S^{a}_{bc}$ for non-static, spherically
symmetric spacetimes, using the space-time metric \eqref{metrica2b}
and the pulled-back material metric \eqref{pulledbk} can be written
as:
\begin{align*}
S^{r}_{rr}&=\frac{f'}{f}-\frac{\bar{b}'}{2\bar{b}}\\
S^{\theta}_{\theta r}&=\frac{f'}{f}+\frac{1}{r}-\frac{\bar{Y}'}{\bar{Y}}\\
S^{\phi}_{\phi r}&=\frac{f'}{f}+\frac{1}{r}-\frac{\bar{Y}'}{\bar{Y}}\\
S^{r}_{\theta
\theta}&=-\frac{f'r^{2}}{f}-\frac{1}{r}+\frac{\bar{Y}'\bar{Y}}{\bar{b}}\\
S^{r}_{\phi\phi}&=-\frac{r^{2}f'\sin^{2}\theta}{f}-\frac{r\sin^{2}\theta}{1}+\frac{\bar{Y}'\bar{Y}\sin^{2}\theta}{\bar{b}}.
\end{align*}

In this case, the pulled-back material metric $k_{ab}$ is \be
k_{ab}=n_{1}^{2}\hspace{0.6mm}x_{a}x_{b}+n_{2}^{2}\hspace{0.6mm}(y_{a}y_{b}+\hspace{0.6mm}z_{a}z_{b}).
\ee Here, $x,y, z$ are eigenvectors of $k^{a}_{b}$ with eigenvalues
$n_{1}^{2}$ and $n_{2}^{2}=n_3^2 $ which depend on $t$ and $r$
according to
\begin{align} n_{1}^{2}&
=f^{2}\frac{1}{b}
\quad & n_{2}^{2}& =n_{3}^{2}=f^{2}\frac{r^{2}}{Y^{2}}.
\end{align}
The \edt \ can be decomposed along the directions determined by the
eigenvectors of $k^{a}_{b}$ as follows \be
S^a_{\hspace{0.15cm}bc}=\underset{\hspace{-0.3cm}1}{M_{bc}}x^a+
\underset{\hspace{-0.3cm}2}{M_{bc}}y^a+\underset{\hspace{-0.3cm}3}{M_{bc}}z^a
,\ee where $\underset{i}{M}$, $i=1,2,3$ are second order, symmetric
tensors (see \cite{VB}). It should be noticed that the eigenvectors

Here we determine the eigenvectors and eigenvalues of these tensors,
complementing the results obtained in \cite{VB} for the static case,
the result being summarized in Tables 1,2 and 3:
\\
\begin{center}
Table 1 - Eigenvectors and eigenvalues for $\underset{1}{M}$\\
\begin{tabular}{|c|c|}
\hline Eigenvectors & Eigenvalues\\
\hline $x$ & $\mu_{1}=\frac{f'}{\sqrt{\bar{b}}f}-\frac{\bar{b}'}{2\bar{b}\sqrt{\bar{b}}}$\\
$y$ &
$\mu_{2}=\frac{\bar{Y}'}{\sqrt{\bar{b}}\bar{Y}}-\frac{r^{2}f'\sqrt{\bar{b}}}{\bar{Y}^{2}f}-\frac{r\sqrt{\bar{b}}}{\bar{Y}^{2}}$\\
$z$ &
$\mu_{3}=\mu_{2}$\\
\hline
\end{tabular}
\end{center}

\begin{center}
Table 2 - Eigenvectors and eigenvalues for $\underset{2}{M}$\\
\begin{tabular}{|c|c|}
\hline Eigenvectors & Eigenvalues\\
\hline $x+y$ & $\mu_{4}=\frac{f'}
{\sqrt{\bar{b}}f}+\frac{1}{\sqrt{\bar{b}}r}-\frac{\bar{Y}'}{\bar{Y}\sqrt{\bar{b}}}$\\
$x-y$ & $\mu_{5}=-\mu_{4}$\\
$z$ & $\mu_{6}=0$\\
\hline
\end{tabular}
\end{center}

\begin{center}
Table 3 - Eigenvectors and eigenvalues for $\underset{3}{M}$\\
\begin{tabular}{|c|c|}
\hline Eigenvectors & Eigenvalues\\
\hline $x+z$ & $\mu_{7}=\mu_{4}$\\
$x-z$ &
$\mu_{8}=-\mu_{4}$\\
$y$ &
$\mu_{9}=0$\\
\hline
\end{tabular}
\end{center}
Therefore, the canonical forms for the three tensors
$\underset{i}{M}$ are: \be \begin{split}
\underset{\hspace{-0.3cm}1}{M_{bc}}&=\mu_{1}x_{b}x_{c}+\mu_{2}(y_{b}y_{c}+z_{b}z_{c})\\
\underset{\hspace{-0.3cm}2}{M_{bc}}&=2\mu_{4}(x_{b}y_{c}+y_{b}x_{c})\\
\underset{\hspace{-0.3cm}3}{M_{bc}}&=2\mu_{4}(x_{b}z_{c}+z_{b}x_{c}).
\end{split}\ee Although the eigenvalues are different from the ones obtained
in the static case, the eigenvectors of the above tensors are the
same for the static and non-static case.

\section{Conclusions}
In this paper we have considered spherically symmetric spacetimes with elastic material content .
We started considering the symmetries these sapcetimes posses in order to exploit their
consequences on physics for elastic spacetime configurations. By doing this, we
have generalized previous work done by \cite{Magli1} and \cite{M1}
for non-static spherically symmetric configurations, where only flat
material metrics were considered. In fact, we have shown that all
material metrics compatible with a given spacetime are conformally
related and, moreover, are conformally flat. Next we have used
comoving coordinates to relate the EFEs with quantities
characterizing elasticity properties (constitutive equation,
material and energy density, eigenvalues of the pulled back material
metric) as well as the conformal factor referred to above. The case in which the velocity
of the matter is shearfree has been considered in detail, giving the necessary and
sufficient condition for a constitutive equation be admitted; further, we have provided three examples
of static and non-static shearfree solutions. For non-static spherically symmetric space-times, the
elasticity difference tensor has been studied, thus extending some previous work for
the static case.

\section*{Acknowledgements}

One of the authors acknowledges financial support from the Spanish
Ministry of Education (MEC) through grant No.: HP2006-0074. Partial
financial support from (MEC) through grant FPA-2007-60220 and from
the ``Govern de les Illes Balears'' is also acknowledged. Further,
this author wishes to express his gratitude for the hospitality at
the Universidade do Minho, where most of this work was done.\\
The other authors acknowledge financial support from (CRUP) through
grant No.: E-89/07 and from FCT and CMAT. They also express their
thanks and gratitude for the hospitability at the Universitat de les
Iles Balears.

\end{document}